\documentclass[amsmath,amssymb,superscriptaddress,aps,prd,reprint,twocolumn,showpacs]{revtex4-1}
\usepackage{graphicx}

\usepackage[colorlinks=true,linkcolor=blue,citecolor=blue,urlcolor=blue]{hyperref}

\newcommand{\dd}{{d}}

\newcommand{\om}{\omega}
\newcommand{\ommax}{\om_{\rm max}}

\begin{document}

\title{Impossibility of superluminal travel in Lorentz violating theories}

\author{Antonin Coutant}
\email{antonin.coutant@th.u-psud.fr}
\affiliation{Laboratoire de Physique Th\'eorique, CNRS UMR 8627, B\^at. 210, Universit\'e Paris-Sud 11, 91405 Orsay Cedex, France}
\author{Stefano Finazzi}
\email{finazzi@science.unitn.it}
\affiliation{INO-CNR BEC Center and Dipartimento di Fisica, Universit\`a di Trento, via Sommarive 14, 38123 Povo-Trento, Italy}
\author{Stefano Liberati}
\email{liberati@sissa.it}
\affiliation{SISSA, Via Bonomea 265, 34136, Trieste, Italy and INFN, Sezione di Trieste}
\author{Renaud Parentani}
\email{renaud.parentani@th.u-psud.fr}
\affiliation{Laboratoire de Physique Th\'eorique, CNRS UMR 8627, B\^at. 210, Universit\'e Paris-Sud 11, 91405 Orsay Cedex, France}

\date{\today}

\begin{abstract}
Warp drives are space-times allowing for superluminal travel. However, they are quantum mechanically unstable because they produce a Hawking-like radiation which is blue shifted at their front wall without any bound.
We reexamine this instability when local Lorentz invariance is violated at ultrahigh energy by dispersion, as in some theories of quantum gravity. Interestingly, even though the ultraviolet divergence is now regulated, warp drives are still unstable. Moreover the type of instability is different whether one uses a subluminal or a superluminal dispersion relation. In the first case, a black-hole laser yields an exponential amplification of the emitted flux whereas, in the second, infrared effects produce a linear growth of that flux.
These results suggest that chronology could still be protected when violating Lorentz invariance.
\end{abstract}

%\keywords{warp drive, Hawking radiation, horizons
% \keywords{warp drive, superluminal travel, Lorentz violating theories, Hawking effect, black hole laser}
\maketitle

\section{Introduction}

Warp drives~\cite{alcubierre} allow, at least theoretically, to travel at arbitrary high superluminal speeds and consequently to travel in time~\cite{everett}.
However, besides the fact that they require matter distributions violating positive energy conditions~\cite{pfenningford,broeck,lobovisser2004a,lobovisser2004b}, 
they are quantum mechanically unstable because they possess a white hole horizon and a Cauchy horizon on which the renormalized stress-energy tensor blows up exponentially~\cite{BFL}.
In this paper, we reexamine the question of their stability when postulating that Lorenz invariance is broken at ultrahigh energy.
In this we were motivated by the fact that nonlinear dispersion relations remove Cauchy horizons and regulate the fluxes emitted by white holes~\cite{MacherRP1}.

A superluminal warp drive metric describes a bubble containing an almost flat region, moving at some constant speed $v_0 > c$
within an asymptotically flat space-time:
\begin{equation}\label{eq:3Dalcubierre}
 ds^2=-c^2 dt^2+\left[dx-v(r)dt\right]^2+dy^2+dz^2,
\end{equation}
where $r\equiv \sqrt{(x-v_0 t)^2+y^2+z^2}$ is the distance from the center of the bubble.
Here $v=v_0 f(r)$, with $f$ a smooth function satisfying $f(0)=1$ and $f(r) \to 0$ for $r \to \infty$.
Along the direction of motion, the backward and forward locii where $v(r) = c$ behave, respectively, as a future (black) and past (white) event horizon~\cite{hiscock}.
In fact, the Hawking flux emitted by the black horizon accumulates on the white horizon while being unboundlessly blueshifted.
However, since the whole analysis rests on relativistic quantum field theory, one should examine whether the warp drive's instability is peculiar to the local Lorentz symmetry.
Although current observations constrain to ultrahigh energy a possible breaking of that symmetry~\cite{Liberati:2009pf}, one cannot exclude this possibility which has been suggested by theoretical  investigations~\cite{Gambini:1998it,Jacobson-Mattingly,Horava}.
%as well as hinted by the recent OPERA measurement~\cite{Opera} (see e.g.~\cite{Maccione:2011fr} for a critical discussion).

\begin{figure}
 \includegraphics[width=0.99\columnwidth]{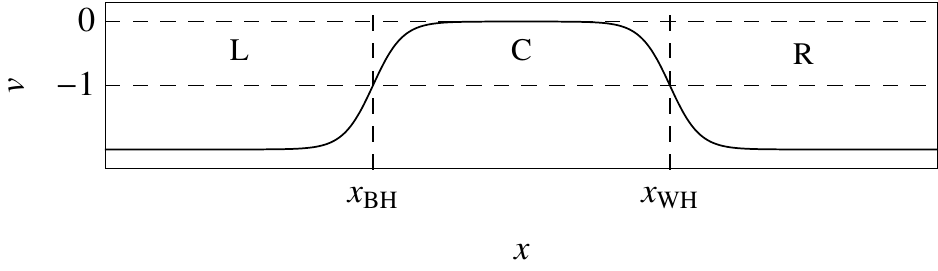}
 \caption{Velocity profile for a right-going warp drive in the Painlev\'e-Gullstrand~\cite{livrev} coordinates of Eq.~\eqref{eq:pg}. Two {\it superluminal} asymptotic regions $L$ and $R$ are separated by a black and a white horizon from a compact internal {\it subluminal} region $C$.
The Killing field $\partial_t$ is spacelike in $L$ and $R$, lightlike on both horizons, and timelike in $C$.}
 \label{fig:velocity}
\end{figure}

For the sake of simplicity we work in $1+1$ dimensions and ignore the transverse directions $y$ and $z$.
Defining a new spatial coordinate $X=x-v_0 t$, the metric becomes
\begin{equation}\label{eq:pg}
 ds^2=-c^2 dt^2+\left[dX-V(X)dt\right]^2,
\end{equation}
where $V(X) = v_0(f(X) - 1)$ is negative. In this space-time, $\partial_t$ is a globally defined Killing vector field whose norm is given by $c^2 - V^2$: it is timelike within the bubble, its norm vanishes on the two horizons, and it is spacelike outside.
One thus gets three regions: $L$, $C$, and $R$ (see Fig.~\ref{fig:velocity}), separated by two horizons $x_{\rm BH}<0<x_{\rm WH}$, which are, respectively, the black and the white one.

\begin{figure}[b]
\includegraphics[width=0.48\columnwidth]{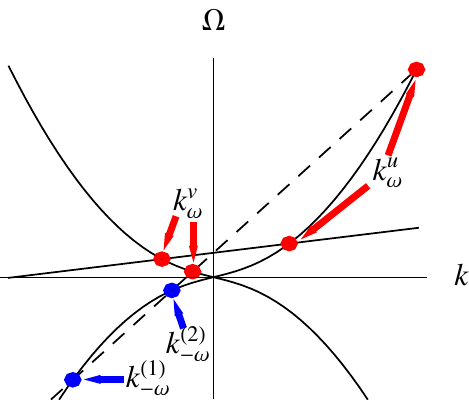}
\includegraphics[width=0.48\columnwidth]{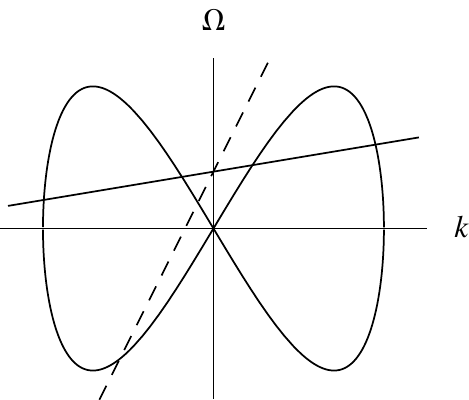}
 \caption{(color online). Graphical solution of Eq.~\eqref{eq:dispersion} for super (left panel), and subluminal dispersion (right panel). In both panels, the straight lines represent $\omega-V k$ for $|V|<1$ (solid) and $|V|>1$ (dashed). 
The curved lines represents $\pm \Omega_\pm(k)$.
On the left, red (blue) dots in the upper (lower) half plane refer to roots with positive (negative) $\Omega_+$ %red (blue) dots refer to roots with positive (negative) $\Omega_+$ 
which correspond to  positive (negative) norm modes.}
 \label{fig:dispersion}
\end{figure}

We now consider a massless scalar field with a quartic dispersion relation.
In covariant terms, its action reads 
\begin{equation}
  S_\pm=\frac{1}{2}\int\! d^2 x
\sqrt{-g}\left[g^{\mu\nu}\partial_{\mu}\phi
\partial_{\nu}\phi \pm \frac{(h^{\mu\nu}\partial_{\mu}\partial_{\nu}\phi)^2}{\Lambda^2}\right],
\label{action}
\end{equation}
where $h^{\mu\nu}=g^{\mu\nu}+u^\mu u^\nu$ is the spatial metric in the direction orthogonal to a unit timelike vector field $u^\mu$.
This extra background field specifies the preferred frame used to implement the dispersion relation~\cite{Tedprd96}.
In the present settings $u^\mu$ should be given from the outset, while in condensed matter the preferred frame is fixed by the fluid flow~\cite{Unruh95}. Inspired by this analogy, we choose $u^\mu$ to be $(1,V)$ in the $t,X$ frame, i.e.~stationarity is preserved.
Then the {\it aether} flow is geodesic and it is asymptotically at rest in the $t,x$ frame of Eq.~(\ref{eq:3Dalcubierre}).
The sign $\pm$ in Eq.~(\ref{action}) holds for superluminal and subluminal dispersion, respectively.
Using Eq.~\eqref{eq:pg} and  $u^\mu=(1,V)$, the wave equation is
\begin{equation}
 \left[\left(\partial_t+\partial_X V\right)\left(\partial_t+V\partial_X \right)-\partial_X^2\pm\frac{1}{\Lambda^2}\partial_X^4\right]\phi=0. 
\label{modeequation}
\end{equation}
Because of stationarity, the field can be decomposed in stationary modes $\phi = \int d\omega\, e^{- i \omega t }\phi_\omega$, where $\omega$ is the conserved (Killing) frequency. Correspondingly, at fixed $\omega$ the dispersion relation reads
\begin{equation}\label{eq:dispersion}
(\omega-V k_\om)^2=k_\om^2\pm\frac{k_\om^4}{\Lambda^2}\equiv\Omega_\pm^2,
\end{equation}
where $k_\om(X)$ is the spatial wave vector, and $\Omega$ the {\it comoving} frequency, i.e.~the frequency in the aether frame. The graphical solution of Eq.~\eqref{eq:dispersion} is plotted in Fig.~\ref{fig:dispersion}.
For superluminal dispersion and $|V|<1$, there are two real roots ($k^v_\om$, $k^u_\om$) describing left- and right-going waves ($\phi^{v}_{\omega}$, $\phi^{u}_{\omega}$), and two complex ones ($k^\uparrow_\om$, $k^\downarrow_\om$) describing a spatially growing and decaying mode ($\phi^\uparrow_{\omega}$, $\phi^\downarrow_{\omega}$).
For $|V|>1$, there is a cutoff frequency $\ommax$~\cite{MacherRP1} below which the complex roots turn into real ones ($k^{(1)}_\om$, $k^{(2)}_\om$) with negative $\Omega$.
Correspondingly there exist two additional propagating waves $(\phi^{(1)}_{-\omega})^*$, $(\phi^{(2)}_{-\omega})^*$ with negative norm.
When the dispersion relation is subluminal,
%  the negative norm modes 
the two extra roots correspond to positive norm modes that
are trapped in the region with $|V|<1$.

\section{Superluminal dispersion relation}
\begin{figure}
\includegraphics[width=\columnwidth]{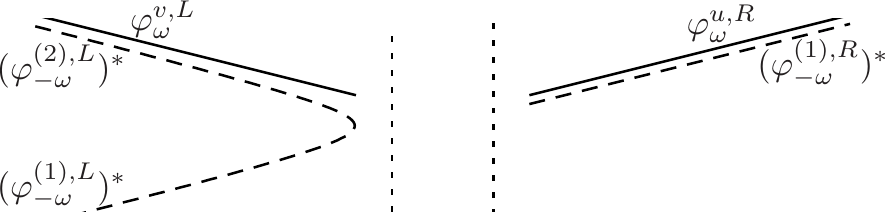}
 \caption{Asymptotic decomposition in plane waves $\varphi^{(i),  L/R}_\om$ of the {\it in} mode $(\phi^{(1),\rm in}_{-\om})^*$. Note that only $\varphi^{(1),  L}_{-\om}$ has group velocity directed toward the horizons, with wavevector $k_\om^{(1)}$.}
\label{fig:mode}
\end{figure}

% \emph{Superluminal dispersion relation}.---
In a geometry with two infinite asymptotic ``superluminal'' regions, for each $\om<\ommax$, 4 asymptotically bounded modes~\cite{MacherRP1} can be defined.
Moreover, by examining their asymptotic behavior, an \emph{in} and an \emph{out} bases can be defined by the standard procedure: each \emph{in} mode $\phi^{(i),\rm in}_{\om}$ (\emph{out} mode $\phi^{(i),\rm out}_{\om}$) possesses a single asymptotic branch $\varphi^{(i),  L/R}_\om$ carrying unit current and with group velocity directed towards region $C$ (from $C$ to $\infty$).
This is exemplified in Fig.~\ref{fig:mode} using the \emph{in} mode $\phi^{(1),\rm in}_{-\om}$.

When the dispersive scale and the horizon surface gravity $\kappa$ are well separated ($\om \sim \kappa \ll \Lambda$), the left-going mode does not significantly mix with the other three modes, all defined on the right-going branch of Eq.~\eqref{eq:dispersion}~\cite{ACRPSF}, as we checked by numerical techniques based on~\cite{MacherRP1}.
Thus, the {\it in}-{\it out} scattering matrix is effectively a $3\times3$ matrix
\begin{equation}
\begin{pmatrix}
 \phi^{u,\rm in }_{\om}\\
 \left(\phi^{(1),\rm in }_{-\om}\right)^*\\
 \left(\phi^{(2),\rm in }_{-\om}\right)^*\\
\end{pmatrix}
=\begin{pmatrix}
 \alpha_{\om}^u & \beta^{(1)}_{-\om} & \beta^{(2)}_{-\om}\\ 
 \beta^{(1)}_{\om} & \alpha_{-\om}^{(1)} & A_{-\om}\\ 
 \beta^{(2)}_{\om} & \tilde A_{-\om} & \alpha_{-\om}^{(2)} 
\end{pmatrix}
\begin{pmatrix}
 \phi^{u,\rm out}_{\om}\\
 \left(\phi^{(1),\rm out}_{-\om}\right)^*\\
 \left(\phi^{(2),\rm out}_{-\om}\right)^*\\
\end{pmatrix}.
\label{eq:bogo4x4}
\end{equation}
Given that the two $(\phi^{(i)}_{-\om})^*$ have negative norms, the matrix coefficients satisfy normalizations conditions such as
\begin{equation} 
| \alpha^{u}_{\om} |^2-| \beta^{(1)}_{\om}|^2 - | \beta^{(2)}_{\om} |^2=1.\label{current}
\end{equation}
When working in the $in$-vacuum, the state without incoming particles, the mean occupation numbers of outgoing particles with negative frequency are $\bar n^{(i)}_{-\om}= |\beta^{(i)}_{-\om} |^2$, whereas that  with positive frequency is $\bar n^{u}_{\om}=\bar n^{(1)}_{-\om}+ \bar n^{(2)}_{-\om}$, by energy conservation.
That is, pair production here occurs through a two-channel Hawking-like mechanism.
To approximately compute the coefficients of Eq.~\eqref{eq:bogo4x4} in the regime $\kappa\ll \Lambda$ we use connection formula techniques. 
We first decompose $\phi_\om$ in both asymptotic regions $L$ and $R$ as a sum of plane waves:
\begin{eqnarray}
\phi_\om &=& L_\om^u \, \varphi_\om^{u,L}+ L^{(1)}_{\om}  (\varphi^{(1),L}_{-\om})^* + L^{(2)}_{\om} (\varphi^{(2),L}_{-\om})^*,
\nonumber\\
\phi_\om &=& R_\om^u \, \varphi_\om^{u,R} + R^{(1)}_{\om} (\varphi^{(1),R}_{-\om})^* + R^{(2)}_{\om} (\varphi^{(2),R}_{-\om})^*.
\end{eqnarray}
The coefficients are connected by
\begin{equation}\label{eq:connection}
 \begin{pmatrix}R_\om^u \\ R_\om^{(1)} \\ R_\om^{(2)} \end{pmatrix} = U_{\rm WH} \cdot U_{\rm HJ} \cdot U_{\rm BH}^{-1} \cdot \begin{pmatrix}L_\om^u \\ L_\om^{(1)} \\ L_\om^{(2)} \end{pmatrix},
\end{equation}
where $U_{\rm BH}$ and $U_{\rm WH}$, respectively, describe the {\it off-shell} scattering on the two horizons~\cite{ACRPSF} and $U_{\rm HJ}$ describes the WKB propagation from one horizon to the other, i.e. it is diagonal and contains the exponential of $i S_\om^a= i\int{dx}\,k^a_\om(x)$, where $k^a_\om$ is $k^u_\om$, $k^\uparrow_\om$, or $k^\downarrow_\om$.
By {\it off-shell} we mean that these three matrices are not restricted to the two modes that govern the asymptotic scattering on each horizon considered separately, that is the growing mode is here kept in the mode mixing.
In fact, since $k^\uparrow$ has negative imaginary part, $e^{iS^\uparrow_\omega}$ is exponentially large. Simple WKB algebra shows that it grows as $e^{\Lambda \Delta}$ where $\Delta$ is the distance between the two horizons.
Concomitantly, since $k^\downarrow=k^{\uparrow*}$, $e^{iS^\downarrow_\omega}$ is exponentially small.

Let us show with an example how to determine the coefficients of Eq.~\eqref{eq:bogo4x4}.
The globally defined $\phi^{(1),\rm in}_{-\om}$ is constructed by imposing that the asymptotic amplitudes of the two incoming branches $\varphi^{u, L}_\om, \varphi^{(2), R}_{-\om}$ both vanish, see Fig.~\ref{fig:mode}.
Therefore, the three outgoing amplitudes are given by the second row of the matrix of Eq.~\eqref{eq:bogo4x4}. Moreover, using Eq.~\eqref{eq:connection}, these coefficients correspond to $(R_\om^u,R_\om^{(1)},R_\om^{(2)})=(\beta_\om^{(1)} , \alpha_{-\om}^{(1)} , 0) $ and $(L_\om^u , L_\om^{(1)} , L_\om^{(2)}) =(0 , 1 , A_{-\om})$. 
Solving the resulting system, we obtain
\begin{equation}
\begin{aligned}
 \beta^{(1)}_{\om}&=\tilde\beta_\om^{\rm BH} \times e^{iS^u_\om} \times \alpha_\om^{\rm WH}+O(e^{iS_\om^\downarrow}),
\\
 \alpha_{-\om}^{(1)}&= - \tilde \beta_\om^{\rm BH} \times e^{iS^u_\omega} \times \beta_\om^{\rm WH} +O(e^{iS_\om^\downarrow}),\label{3Bogo}
\\
 A_{-\om}&=\tilde\alpha_\om^{\rm BH},
\end{aligned}
\end{equation}
where the $\alpha$'s and $\beta$'s in the above are the standard Bogoliubov coefficients for black and white holes that encode the thermal Hawking radiation~\cite{Primer}.
By a similar analysis of other modes, all coefficients of Eq.~\eqref{eq:bogo4x4} can be computed.
Although the nonpositive definite conservation law of  Eq.~(\ref{current}) does not bound these coefficients, the exponentially large factor in $e^{\Lambda \Delta}$ cancels out from all of them.
As a consequence, as can be seen in Eq.~\eqref{3Bogo}, the leading term is, up to some phase coming from the WKB propagation in region $C$, given by the Bogoliubov coefficients of $U_{\rm BH}$ and $U_{\rm WH}$~\cite{ACRPSF}. 
In the first two lines, one finds a product of two coefficients because the associated semiclassical trajectory passes through both horizons. Instead, in the third line 
only one coefficient is found because there is only a reflection on the black horizon.

To identify the physical consequences of the pair creation encoded in Eq.~\eqref{eq:bogo4x4}, we compute the expectation value of the stress-energy tensor
\begin{equation}
 T_{\mu\nu} \equiv \frac{2}{\sqrt{-g}}\frac{\delta S_+}{\delta g^{\mu\nu}}=T_{\mu\nu}^{(0)}+T_{\mu\nu}^{(\Lambda)},
\end{equation}
where $T_{\mu\nu}^{(0)}$ is the standard relativistic expression and $T_{\mu\nu}^{(\Lambda)}$ arises from the Lorentz violating term of Eq.~\eqref{action}: 
\begin{multline}
 T_{\mu\nu}^{(\Lambda)} =\frac{1}{\Lambda^2}
 \left[h^{\alpha\beta}\left(\phi_{,\alpha\beta}\phi_{,\mu\nu}+\phi_{,\mu\nu}\phi_{,\alpha\beta}\right)
\right.\\\left.
-\frac{1}{2}\left(h^{\alpha\beta}\phi_{,\alpha\beta}\right)^2g_{\mu\nu}\right].
\end{multline}
In the asymptotic region on the right of the white horizon, the field can be expanded as the superposition of the two right-going modes $\phi_\omega^{u,\rm out}$ and $\phi_{-\omega}^{(1),\rm out}$, see Fig.~\ref{fig:mode},
\begin{equation}
 \phi=\int\dd\omega\left[\phi_\omega^{u,\rm out} \hat a_\omega^{u,\rm out}+\phi_{-\omega}^{(1),\rm out} \hat a_{-\omega}^{(1),\rm out}\right]+{\rm h.c.}
\end{equation}
In this region, the geometry is stationary and homogeneous. Hence, the renormalized tensor $T_{\mu\nu}^{\rm ren}$ is obtained by normal ordering the above creation and destruction {\it out}\/ operators.
[The fact that there exist negative frequency asymptotic particles causes no problem in this respect.
In fact, all asymptotic excitations have a positive comoving frequency $\Omega$ of Eq.~\eqref{eq:dispersion}.]
Imposing that the initial state is vacuum, using Eq.~\eqref{eq:bogo4x4}, it is straightforward to compute $\langle 0_{\rm in}|T_{\mu\nu}^{\rm ren}|0_{\rm in}\rangle$.
The final expression contains an integral over $\omega$ of a sum of terms, each being the product of two modes $\phi_\omega^{u,\rm out}, \phi_{-\omega}^{(1),\rm out}$ and two coefficients of Eq.~\eqref{eq:bogo4x4}.
We do not need the exact expression because we only consider possible divergences.
When $\omega>\omega_{\rm max}$ there are no negative frequency modes. Hence, the $\beta$ coefficients of Eq.~\eqref{eq:bogo4x4} vanish for $\om > \omega_{\rm max}$, and the stress-energy tensor cannot have ultraviolet divergences.
Therefore, the only possible divergence can be found for $\omega\to0$.

In each term of $T_{\mu\nu}^{(0)}$, there are two derivatives with respect to $t$ or $x$, leading to two powers of $\omega$, $k_\omega^{(u)}$, or $k_\omega^{(1)}$. 
Analogously in $T_{\mu\nu}^{(\Lambda)}$, there are 4 powers of these.
Now, from Fig.~\ref{fig:dispersion} we see that the wavenumbers $k_\omega^{(u)}, k_\omega^{(1)}$ do not vanish for $\omega\to0$ in $L$. Rather, they go to constant opposite values, that we call $k_0$ and $-k_0$, respectively. Interestingly, these modes played a key role in~\cite{CarlosMayo} when studying the fluxes emitted by a white hole flow in an atomic Bose condensate. They are also associated with the macroscopic {\it undulation} observed in the experiments~\cite{Rousseaux,Silke} and theoretically described in Sec.~III.B.3 of~\cite{ACRPSF}.
Finally, the terms containing only spatial derivatives will not be suppressed for $\omega\to0$.
The leading terms in $\langle 0_{\rm in}|T_{\mu\nu}^{(0), \, \rm ren}|0_{\rm in}\rangle$ are thus proportional to
\begin{equation}
\frac{k_0^2}{{4\pi\Omega(k_0)}\, v_{g0}}\, \int\dd\omega \left[\bar n^{(u)}_\om +\bar n^{(1)}_{-\omega}\right].
\label{leading}
\end{equation}
The above integral gives the integrated mean occupation number of the two {\it out}\/ species, and $v_{g0}$ is their asymptotic group velocity in the $t,X$ frame.
The leading terms of $\langle 0_{\rm in}|T_{\mu\nu}^{(\Lambda), \, \rm ren}|0_{\rm in}\rangle$ are proportional to Eq.~\eqref{leading} up to an extra factor of $k_0^2/\Lambda^2$.
Since $k_0 = \Lambda\sqrt{v_0^2-1} $, one finds that the $(0)$ and $(\Lambda)$ components of the stress-energy tensor yield typically the same contribution.

The key result comes from the fact that $|\beta^{(1)}_\omega|^2$
diverges as $1/\omega^2$ for $\omega\to 0$, being the product of $|\beta^{\rm BH}_\omega|^2\sim 1/\omega$ and $|\beta^{\rm WH}_\omega|^2\sim 1/\omega$.
(This infrared behavior has been validated by numerical analysis.)
Then, if the warp drive is created at some time, after some transient period the geometry will become stationary and the outgoing fluxes will reach their stationary values.
However, only frequencies $\omega>1/T$ will contribute after a lapse of time $T$.
This provides an infrared cutoff to the integral of Eq.~\eqref{leading}, which gives an energy density scaling as
\begin{equation}
{\cal E} \propto \Lambda \int_{1/T}\dd\omega \left[\bar n^{(u)}_\om +\bar n^{(1)}_{-\omega}\right]\propto \Lambda \kappa^2 T.
\end{equation}
That is, there is an infrared divergence that leads to a linear growth of the energy density.
This result can be understood from the findings of~\cite{CarlosMayo}: the BH radiation emitted towards the WH horizon stimulates the latter as if a thermal distribution were initially present. In that case, it was also found that the observable (the density correlation function) increased {linearly} in $T$. 

% Using quantum inequalities~\cite{QI}, it was argued~\cite{BFL} that $\kappa$ must be of the order of the Planck scale (say $\kappa\lesssim 10^{-2}t_{\rm P}^{-1}$, where $t_{\rm p}$ is the Planck time), which implies that the growth rate is also of that order (unless $\Lambda$ is very different from that scale). In the presence of superluminal dispersion, warp drives are therefore unstable on a short scale.

Using quantum inequalities~\cite{QI}, it was argued~\cite{BFL} that
%  $\kappa$ must be of the order of the Planck scale (say 
$\kappa\lesssim 10^{-2}t_{\rm P}^{-1}$, where $t_{\rm P}$ is the Planck time, which implies that the growth rate is of the same order (unless $\Lambda$ is very different from $t_{\rm P}^{-1}$). In the presence of superluminal dispersion, warp drives are thus unstable on a short scale.

\section{Subluminal dispersion relation}
% \emph{Subluminal dispersion relation}.---
In this case, the warp drive system is also unstable. 
Indeed, the positive %negative 
norm modes $\phi_{\om}^{(1)}$ and $\phi_{\om}^{(2)}$ are now trapped between the two horizons in region $C$.
Hence, they bounce back and forth and this induces an exponentially growing amplification of the emitted radiation. 
As a result, the asymptotic fluxes, and therefore $\langle 0_{\rm in}|T_{\mu\nu}^{\rm ren}|0_{\rm in}\rangle$, grow exponentially in time. 
This phenomenon is the subluminal version of the black-hole laser effect~\cite{TJlaser}.
This dynamical instability is described by complex frequency eigenmodes that are asymptotically bounded~\cite{CP}.
In the original version, the analysis was performed  with a superluminal dispersion relation and with a metric like that of Eq.~\eqref{eq:pg} but with $|V| < 1$ at $\pm \infty$ and $|V| > 1 $ in the central region (see added note).
However, there is a precise symmetry between that case and the present one~\cite{ACRPSF}.
It basically consists in changing both the sign of the dispersion relation and that of $V + 1$, i.e. interchanging sub and superluminal regions.
This symmetry allows to infer that the complex frequencies governing the laser instability will share the same characteristics as in~\cite{CP}.
In particular, the growth rate of the most unstable mode, which is a non trivial function of the surface gravity of each horizon and their separation $\Delta$, can be read of from the expressions of that work.

\section{Final remarks}
% \emph{Final remarks}.---
In this paper, we generalized~\cite{BFL} by showing that superluminal warp drives are unstable even when local Lorentz invariance is broken at very high energies.
When the dispersion relation is subluminal, the horizons act as a resonant cavity producing a dynamical black-hole laser instability~\cite{TJlaser}. When it is superluminal, instead, the emitted flux grows linearly in time due to infrared effects.

To conclude, we note that in~\cite{everett} it was shown that close timelike curves can be obtained by combining several warp drives.
Our results, together with those of~\cite{BFL}, weaken that possibility because isolated warp drives are unstable irrespectively of the features of the dispersion relation in the ultraviolet regime.
As a consequence, whereas former attempts to tackle the issue of chronology protection deeply relied on local Lorentz invariance~\cite{KRW}, the present result suggests that this conjecture may be valid also for quantum field theories violating Lorentz invariance in the ultraviolet sector.
% \\

% \begin{center}
\subsection*{Added note}
% \end{center}
In fact, in the presence of superluminal dispersion, as it is the case in Ho{\v r}ava gravity~\cite{Horava}, all black holes could be dynamically unstable. This conjecture is based on the observation (see Fig.~10 in~\cite{broad} and the associated discussion) that any significant scattering in the region where $|V| > 1 $ induces signs of instability. It is therefore important to test this conjecture by a linear stability analysis based e.g. on the ``improved'' action of~\cite{Blas_action,Blas_Sibiryakov2011}.

\begin{acknowledgments}
 SF has been supported by the Foundational Questions Institute (FQXi) Grant No.~FQXi-MGA-1002.
\end{acknowledgments}

\end{document}